\documentclass[aps,twocolumn,superscriptaddress,prl,amsmath]{revtex4-2}
\usepackage{multirow}
\usepackage[latin9]{inputenc}
\usepackage{epstopdf}
\usepackage{mathrsfs}
\usepackage{txfonts}
\usepackage{amssymb}
\usepackage{graphicx,subfigure,float}
\usepackage{dcolumn}
\usepackage{bbm}
\usepackage{bm}
\usepackage{color}
\usepackage[colorlinks, linkcolor=blue, citecolor=blue, urlcolor=blue]{hyperref}
\usepackage{lipsum}
\usepackage{makecell}
\usepackage{gensymb}
\usepackage{pifont}
\usepackage{placeins}
\begin{document}
\title{Layer Edelstein Effect}
  \author{Binchang Zhou}
\affiliation{Key Laboratory of Low Dimensional Materials and Application Technology of Ministry of
Education, School of Materials Science and Engineering, Xiangtan University, Xiangtan 411105, China}
  \author{Pan Zhou}
 \email{zhoupan71234@xtu.edu.cn}
\affiliation{Key Laboratory of Low Dimensional Materials and Application Technology of Ministry of
Education, School of Materials Science and Engineering, Xiangtan University, Xiangtan 411105, China}
\affiliation{Hunan Provincial Key laboratory of Thin Film Materials and Devices, School of Materials Science
and Engineering, Xiangtan University, Xiangtan 411105, China}
\author{Baoru Pan}
\affiliation{Hunan Provincial Key laboratory of Thin Film Materials and Devices, School of Materials Science
and Engineering, Xiangtan University, Xiangtan 411105, China}
\author{Yuzhong Hu}
\affiliation{Hunan Provincial Key laboratory of Thin Film Materials and Devices, School of Materials Science
and Engineering, Xiangtan University, Xiangtan 411105, China}
\author{Songmin Liu}
\affiliation{Hunan Provincial Key laboratory of Thin Film Materials and Devices, School of Materials Science
and Engineering, Xiangtan University, Xiangtan 411105, China}
\author{Lizhong Sun}
 \email{lzsun@xtu.edu.cn}
\affiliation{Key Laboratory of Low Dimensional Materials and Application Technology of Ministry of
Education, School of Materials Science and Engineering, Xiangtan University, Xiangtan 411105, China}

\date{\today}
\begin{abstract}
Electrical control of magnetism represents a fundamental route toward next-generation spintronic functionalities. In this Letter, we introduce a universal current-induced spin phenomenon in bilayer systems, termed the layer Edelstein effect (LEE), which serves as the natural counterpart of the layer Hall effect in real space. It is defined by the emergence of layer-resolved spin magnetizations with opposite components on the top and bottom layers, driven by an in-plane charge current and controllable by an external electric field. We establish the general existence of the LEE using a minimal bilayer $\bm{k \cdot p}$ theory. By combining symmetry analysis with a general bilayer stacking framework, we derive a model-independent symmetry criterion demonstrating that the LEE is generically allowed in a broad class of nonmagnetic bilayer stacking systems. We further show that the LEE admits two universal manifestations: explicit layer-opposite spin magnetization components mandated directly by symmetry, and components become activated upon symmetry reduction by external electric fields. First-principles calculations on stacked bilayer MoSSe, MoTe$_2$ and WTe$_2$ confirm the predicted effect and illustrate their experimental feasibility. Our work establishes the LEE as a generic symmetry-governed response of bilayer systems, providing a unified conceptual framework for electrically generating and manipulating layer-resolved spin polarization.
\end{abstract}
\maketitle
\textit{Introduction}---The generation and manipulation of magnetism constitute central themes in spintronics and magnetic memory technologies\cite{in1,in2,in3,in4,in5,in6,in7,in8}. Conventional magnetic storage relies on magnetization switching driven by locally generated magnetic fields, which is intrinsically limited by weak coupling efficiency, poor spatial selectivity, and high energy consumption. These limitations have motivated extensive efforts toward electric-field-based control schemes, which offer faster, more energy-efficient routes for manipulating magnetic degrees of freedom\cite{in9,in10,in11,in12,in13,in14,in15,in16,in17}. In this context, current-induced spin responses play a pivotal role by enabling electrical generation of magnetization without invoking magnetic order. Among them, the Edelstein effect represents a prototypical mechanism, whereby an applied electric current induces a nonequilibrium spin polarization in noncentrosymmetric systems\cite{edelstein-effect,REE1,REE2}, as confirmed experimentally\cite{ex-confirm1,ex-confirm2}. Owing to its conceptual simplicity and broad applicability, the Edelstein effect has stimulated a wide range of theoretical and experimental extensions\cite{rashba1,rashba2,rashba3,rashba4,rashba5,rashba6,rashba7,rashba8,rashba9,rashba10,out1,out2,out3,out4,anomalous,
orbit1,orbit2,orbit3,orbit5,nonlinear,nonrel1,nonrel2,nonrel3,inverse-edel,inverse-edel2}.\\
\indent More recently, the emergence of layertronics\cite{lst1,lst2,lst3,lst4,lst5} has established the layer index as an additional internal degree of freedom for controlling electronic responses in low-dimensional systems. In inversion-asymmetric bilayers, this degree of freedom enables layer-resolved transport phenomena, most notably the layer Hall effect\cite{layer-hall1,layer-hall2,layer-hall3,layer-hall4,layer-hall5}, where electrons in the top and bottom layers deflect oppositely while the net charge Hall response vanishes. These observations naturally suggest that current-induced responses need not be uniform across layers, but may acquire opposite or distinct characteristics tied to the layer degree of freedom. These developments naturally raise a fundamental question: can the Edelstein effect be generalized to a layer-resolved form in nonmagnetic bilayers, and if so, can such a layer-dependent current-induced spin polarization be efficiently controlled by an external electric field?\\
\indent In this Letter, we answer this question by introducing the \emph{layer Edelstein effect} (LEE), the natural counterpart of the layer Hall effect in bilayer systems. It is characterized by the generation of opposite spin magnetizations on the two constituent layers under an in-plane current, which can be selectively controlled by an external out-of-plane electric field. First, we establish its existence using a minimal bilayer $\bm{k \cdot p}$ model. By combining symmetry analysis with a general bilayer stacking theory, we further demonstrate that the LEE is a \emph{generic and symmetry-governed phenomenon}, realizable in a wide class of nonmagnetic bilayer stackings, including both centrosymmetric and noncentrosymmetric systems. Remarkably, we show that its emergence is determined solely by the \emph{point-group symmetry of the bilayer}, providing a universal and practical criterion applicable to all 80 layer groups and their stacking configurations. Further first-principles calculations on representative materials of MoSSe, MoTe$_2$ and WTe$_2$ bilayers confirm our predictions.\\
\indent The Edelstein effect describes the generation of a nonequilibrium spin density by an applied electric current in noncentrosymmetric systems\cite{edelstein-effect,REE1,REE2}. Owing to the absence of inversion symmetry, an external current shifts the electronic distribution in momentum space, leading to an imbalance of spin populations and hence a finite spin polarization\cite{nonzero1,nonzero2}. Within the Boltzmann linear transport theory, the induced nonequilibrium spin density is given by
\begin{equation}
\delta S_j = \chi_{ji}\, j_i^{A},
\end{equation}
here \( j_i^{A} \) denotes the applied electric current\cite{tensor_define1,tensor_define2}. The response tensor \( \chi_{ji} \) characterizes the efficiency of current-induced spin polarization and is fully determined by the electronic band structure. Within Boltzmann linear transport theory, it can be expressed as
\begin{equation}
\chi_{ji} = -\frac{\sum_{n\boldsymbol{k}} \langle S \rangle_{n\boldsymbol{k}}^{\,j}\,
v_{n\boldsymbol{k}}^{\,i}\,
\left( \partial f_{n\boldsymbol{k}} / \partial E_{n\boldsymbol{k}} \right)}
{e \sum_{n\boldsymbol{k}} \left( v_{n\boldsymbol{k}}^{\,i} \right)^2
\left( \partial f_{n\boldsymbol{k}} / \partial E_{n\boldsymbol{k}} \right)} ,
\end{equation}
where \( \{i,j\} \in \{x,y,z\} \), \( \langle S \rangle_{n\boldsymbol{k}}^{\,j} \), \( v_{n\boldsymbol{k}}^{\,i} \), and \( E_{n\boldsymbol{k}} \) denote the expectation value of the spin operator, the group velocity, and the energy of the \( n \)-th band at crystal momentum \( \boldsymbol{k} \), respectively, and \( f_{n\boldsymbol{k}} \) is the Fermi--Dirac distribution function. Since the applied electric current is a polar vector while the induced spin magnetization is an axial vector, the corresponding response tensor \( \chi_{ij} \) transforms as a second-rank axial tensor, whose nonvanishing components are constrained by the symmetries of the system\cite{rashba2,book2}. Concretely, according to Neumann's principle, the form of $\chi_{ij}$ is constrained solely by the point-group symmetries of the system. Namely, a layer-group operation $(R|\tau)$ imposes the same tensor constraints as its associated point-group operation $(R|0)$ and fractional translations do not alter the allowed tensor components. The detailed discussion about the explicit constitutive relation is given in Sec.~I of the Supplemental Material (SM)\cite{sup}\nocite{select,1T'-MoSSe,transport,frame1,frame3,frame4,frame5,WTe2,CsGeI,FeSi,
Te-2,quantum1,quantum2,pseudopotentials,PBE,WTe2-lat,paoflow1,paoflow2}.\\
\indent Here we recast the Edelstein effect from a layertronics perspective. For a single layer to exhibit a finite Edelstein response, its symmetry must not include spatial inversion. Otherwise, contributions from symmetry-related states cancel exactly, yielding no net spin polarization\cite{rashba2,tensor_define1}. We now consider an inversion-connected bilayer ($B = L + L'$), as shown in Fig. 1. Although the bilayer as a whole is inversion symmetric and therefore exhibits a vanishing global Edelstein response, a finite current-induced spin magnetization can still emerge locally within each individual layer. Specifically, the spin polarization generated in layer ($L$) is exactly compensated by that in the inverted partner layer ($L'$), resulting in zero net magnetization when both layers are considered together. However, a perpendicular electric field can couple to the layer degree of freedom of the electronic bands, enabling layer-selective control of the low-energy states \cite{electric-layer-coupling1,electric-layer-coupling2}. By tuning the electric field, one can selectively enhance the contribution of a specific layer's bands at the Fermi level, thereby lifting the compensation between ($L$) and ($L'$). An upward electric field induces a spin-up polarization localized in the upper layer, whereas a downward field generates a spin-down polarization localized in the lower layer. This mechanism indicates that we can not only modulate the direction of magnetism in the bilayer via an external electric field, but also control which layer hosts the magnetism.\\
\begin{figure}[t]
\centering
\includegraphics[trim={0.0in 0.0in 0.0in 0.0in},clip,width=\linewidth]{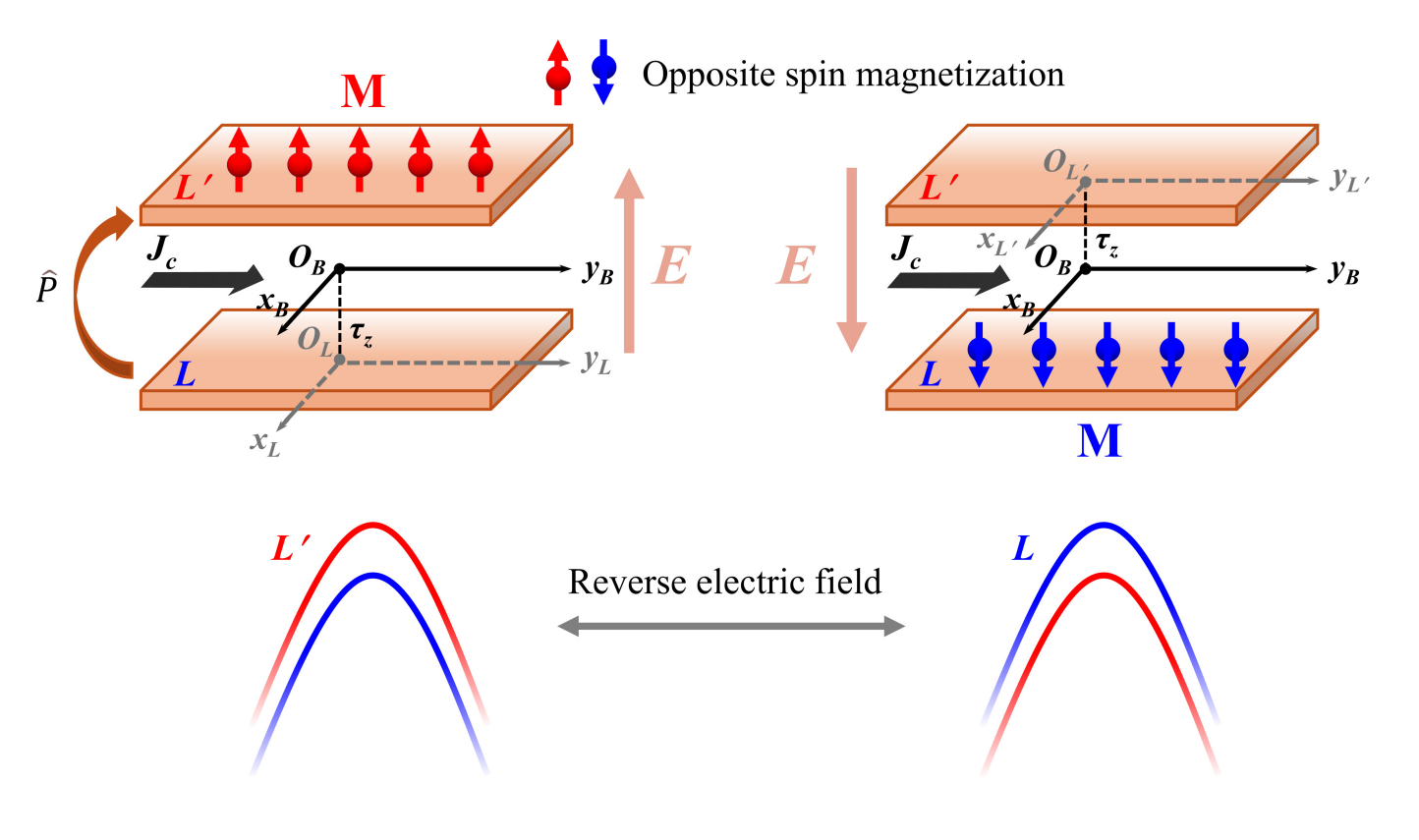}
\caption{Schematic diagram of the LEE. By applying a stacking operator $\hat{P}$, the $L'$ layer can be obtained from the initial single layer $L$. $L$ and $L'$ together form a bilayer system, which uses a bilayer coordinate system labeled $x_BO_By_B$. When we apply an in-plane current $\emph{\textbf{j}}^A$, by changing the direction of the vertical electric field $E$, we can achieve opposite spin magnetization between the layers.}\label{fig1}
\end{figure}
\begin{figure}[t]
\centering
\includegraphics[trim={0.0in 0.0in 0.0in 0.0in},clip,width=\linewidth]{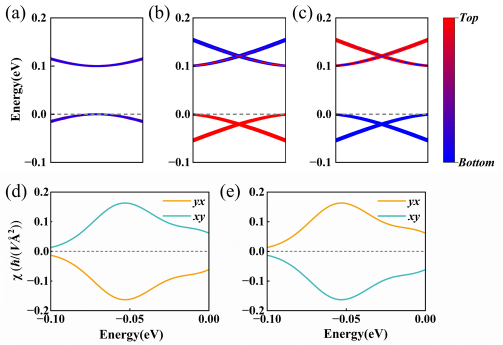}
\caption{Schematic diagram of the $\bm{k \cdot p}$ band structure and corresponding response tensor. (a) $I$ symmetry keeps the top and bottom energy bands degenerate. (b)When a positive out-of-plane electric field is applied, $I$ symmetry is broken, and the top layer's energy band dominate near the Fermi level. (c)When a negative out-of-plane electric field is applied, the bottom layer's energy band dominate near the Fermi level. Calculated $\chi_{xy}$ and $\chi_{yx}$ as a function of the chemical potential when (d) positive and (e) negative electric field is applied.
}\label{fig2}
\end{figure}
\begin{table*}[t]
\caption{Possible bilayer point group obtained by stacking operator $\hat{P}$ and initial layer $L$. Red and blue color represent whether the point group has $Q^-$ operation or not. The labels "\ding{52}" and "\ding{55}" represent whether the LEE is feasible or infeasible, respectively.}
\label{Tab. 1}
\renewcommand{\arraystretch}{1.6}
\setlength{\tabcolsep}{2.2mm}{
\begin{tabular}{ccccccccccccccccc}
\hline  \hline
Bilayer point group & $C_1$ & $C_i$ & $\textcolor[rgb]{0.00,0.00,1.00}{C_2}$ & $\textcolor[rgb]{1.00,0.00,0.00}{C_2}$ & $\textcolor[rgb]{0.00,0.00,1.00}{C_s} $&$\textcolor[rgb]{1.00,0.00,0.00}{C_s}$ & $C_{2h}$ & $D_2$ & $\textcolor[rgb]{0.00,0.00,1.00}{C_{2v}}$ &$\textcolor[rgb]{1.00,0.00,0.00}{C_{2v}} $& $D_{2h}$ & $C_4$ & $S_4$ & $C_{4h}$ & $D_{4}$ & $C_{4v}$\\  \hline
 LEE& \ding{55} & \ding{52} & \ding{55} & \ding{52} & \ding{55} & \ding{52} & \ding{52} & \ding{52} & \ding{55} & \ding{52} & \ding{52} & \ding{55} & \ding{55} & \ding{52} & \ding{52} & \ding{55}  \\
\hline
Bilayer point group & $D_{2d}$ & $D_{4h}$ & $C_{3}$ & $C_{3i}$ & $C_6$ & $D_{3}$ & $C_{3v}$ & $D_{3d}$ & $C_{6}$ & $C_{3h}$ & $C_{6h}$ & $D_{6}$ & $C_{6v}$ & $D_{3h}$ & $D_{6h}$  \\  \hline
LEE & \ding{52} & \ding{52} & \ding{55} & \ding{52} & \ding{55} & \ding{52} & \ding{55} & \ding{52} & \ding{55} & \ding{52} & \ding{52} & \ding{52} & \ding{55} & \ding{52} & \ding{52}  \\
\hline \hline
\end{tabular}}
\end{table*}
\textit{Effective $\textbf{k}\cdot \textbf{p}$ model}---Here, we employ a minimal $\bm{k \cdot p}$ model to demonstrate the layer-dependent Edelstein effect, which we term it as LEE. The low-energy Hamiltonian for the inversion-symmetric bilayer can be written as
\begin{equation}
H(\boldsymbol{k}) = \varepsilon_0 \tau_0\otimes\sigma_0 + t_{\perp} \, \tau_x \otimes \sigma_0 + \Delta \, \tau_z \otimes \sigma_0 + \alpha \, \tau_z \otimes (\sigma_x k_y - \sigma_y k_x),
\end{equation}
where \(\tau_{x,z}\) and \(\sigma_{x,y}\) are Pauli matrices acting on layer and spin subspaces, respectively, \(\tau_0\) and \(\sigma_0\) are identity matrices, \(t_{\perp}\) is the interlayer hopping, \(\Delta\) represents the interlayer potential difference (tunable by an out-of-plane electric field), and \(\alpha\) is the Rashba spin-orbit coupling strength (with opposite signs in the two layers due to the \(\tau_z\) factor). The Rashba term induces a tangential spin texture in momentum space\cite{rashba6}, then supports a finite Edelstein response tensor(\(\chi_{xy} = -\chi_{yx} \neq 0\)), the overall bilayer retains $I$ symmetry when \(\Delta = 0\), leading to opposite spin magnetization in the two layers that cancel exactly, yielding zero net response. Applying a finite \(\Delta = \pm 0.1 t_{\perp}\) breaks $I$ symmetry and shifts spectral weight toward one layer. As shown in Figs.~2(b) and 2(c), this results in asymmetric band dispersions along high-symmetry paths. Consequently, a charge current along the \(x\) (\(y\)) direction induces a net spin magnetization along the \(y\) (\(x\)) direction, with the sign reversing upon reversal of \(\Delta\) [Figs.~2(d) and 2(e)]. This electric-field-tunable, current-induced spin magnetization directly verifies the existence of LEE. It is worth mentioning that interlayer tunneling may, in principle, occur between the two monolayers in a bilayer system. However, its amplitude can be substantially suppressed by combining weak intrinsic interlayer coupling, symmetry-enforced selection rules, enlarged interlayer spacing, and electric-field-induced layer detuning. A detailed discussion is provided in Sec.~II of the SM\cite{sup}. In what follows, we examine the general symmetry conditions of LEE for nonmagnetic bilayers.\\
\indent \textit{Symmetry rules for LEE}---The essential requirement for realizing the LEE is that, under an applied in-plane charge current  $\emph{\textbf{j}}^A$, the induced nonequilibrium spin magnetizations in the upper and lower layers possess at least one pair of corresponding components with opposite signs. In stacked bilayers, the existence and mutual relation of these spin components are fully dictated by the crystalline symmetries of the system\cite{book1,book2,space-group}. Therefore, a symmetry-based formulation provides a natural and rigorous framework for identifying nonmagnetic bilayers capable of hosting the LEE.\\
\indent We consider a bilayer system $B$ constructed from an initial monolayer $L$ and a stacked partner $L'$, generated by the action of a stacking operator $\hat{P}$, i.e., $B = L + \hat{P}L$\cite{stacking1,stacking2}. The stacking operator can be written as $\hat{P} = P\tau_z\tau_o$, where $P$ denotes a point group operation, $\tau_z$ is a translation along the $z$ direction, and $\tau_o$ is an in-plane fractional translation. In general, the symmetry axes of $L$ and $L'$ differ, which leads to directional confusion when comparing the applied current and induced spin magnetization defined in their respective coordinate systems. To avoid this complication, we adopt a unified bilayer coordinate system, in which the two monolayers are symmetrically positioned with respect to the plane $k_z=0$, with $L$ located below and $L'$ above, as shown in Fig.~1(a). In this case, all original monolayer symmetry operations, described by the $3\times 3$ matrix $Q_{\text{rot}}$, are extended to a unified $4\times 4$ form $Q_L$, depending on whether they reverse the $z$ coordinate~\cite{stacking2}. Specifically, we define two sets of operations, $\{Q_L^+\} = \{E, C_z(\theta), M_\beta\}$ and $\{Q_L^-\} = \{C_{2\alpha}, M_z, I, S_{nz}\}$. The operations in $\{Q_L^+\}$ leave the sign of $z$ unchanged, while those in $\{Q_L^-\}$ must be accompanied by an additional $-2\tau_z$ translation along the $z$ direction. Details of the dimensional extension, derivation, and matrix representations are provided in the companion paper, Ref.~\cite{companion_paper}.\\
\indent We focus on bilayers that preserve the in-plane unit cell of the constituent monolayers. In this case, the stacking operation $P$ must belong to the factor group $LG_P=LG_{\mathrm{lat}}/LG_L$, where $LG_L$ is the layer group of the monolayer, and $LG_{\mathrm{lat}}$ is the corresponding symmetry group of the underlying two-dimensional Bravais lattice, which belongs to one of $C_{2h}$, $D_{2h}$, $D_{4h}$, or $D_{6h}$\cite{stacking1,stacking2}. After determining the stacking operator $\hat{P}$, we need to obtain the bilayer point group $LG_B$. Its elements can be divided into two subsets $LG_B^{\mathrm{intra}}$ and $LG_B^{\mathrm{inter}}$. The subset $LG_B^{\mathrm{intra}}$ consists of operations that map each layer onto itself and is given by the intersection $LG_B^{\mathrm{intra}}=LG_L\cap LG_{L'}$, where $LG_{L'}=\hat{P}LG_L\hat{P}^{-1}$ is the layer group of $L'$. The subset $LG_B^{\mathrm{inter}}$ contains operations that exchange the two layers and can be obtained from the intersection $LG_B^{\mathrm{inter}}=\hat{P}LG_L\cap LG_L\hat{P}^{-1}$. Consequently, the full symmetry of the bilayer system is determined as the union set $LG_B=LG_B^{\mathrm{intra}}\cup LG_B^{\mathrm{inter}}$. The results are summarized in Sec.~III of SM\cite{sup}. A more in-depth analysis of the group theory formalism and application are provided in Ref.~\cite{companion_paper}. Having determined the symmetry operations of all stacked bilayers, we next establish the symmetry classification of the LEE. According to the symmetry rules of the LEE, we find that the layer-opposite response can be symmetry-enforced in two fundamentally distinct ways, termed Type-I and Type-II LEE.
\paragraph{(i) Explicit interlayer-opposite components.}
In the first scenario, a tensor component $\chi^L_{ij}$ is symmetry-allowed and already exists in the monolayer $L$, whereas the corresponding component $\chi^B_{ij}$ of the stacked bilayer should be forced to zero by symmetry. This numerical relationship can be expressed as
\begin{equation}
    \chi_{ij}^{L'} = -\chi_{ij}^{L}.
\end{equation}
In this case, a relevant in-plane current can generate equal but opposite nonequilibrium spin polarizations on the two layers. A vertical electric field lowers the bilayer symmetry to a subgroup and the pristine allowed components remain allowed. The role of the vertical electric field in Type-I is then mainly to break the equivalence between the two layers by shifting the layer-resolved electronic states, modifying their occupations, or changing their relative weights near the Fermi level.
\begin{figure}[t]
\centering
\includegraphics[trim={0.0in 0.0in 0.0in 0.0in},clip,width=\linewidth]{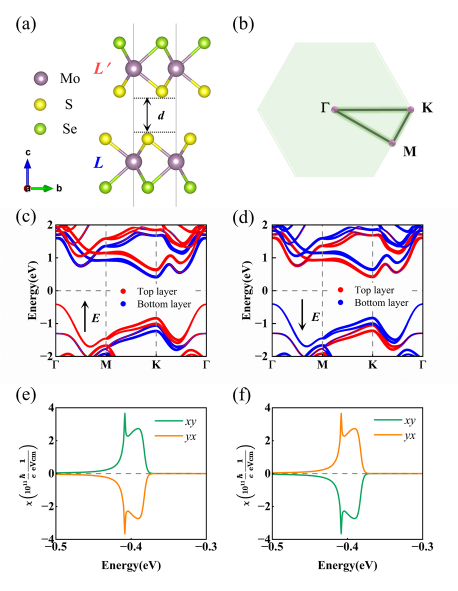}
\caption{(a) For the bilayer system with the stacking of inversion operators, the optimized interlayer spacing is 3.15~\AA, where $L$ and $L'$ represent the bottom and top layer, respectively. (b) Brillouin zone of bilayer MoSSe. (c, d) For the bilayer system under an external electric field of (c) $E = 0.2~\mathrm{eV/\AA}$ and (d) $E = -0.2~\mathrm{eV/\AA}$. (e) The Edelstein tensor calculated corresponding to Fig. (c), where the $Q^+$ operation of the system is retained, corresponds to the component relationship of $\chi_{xy}$=-$\chi_{yx}$. (f) The Edelstein tensor calculated corresponding to Fig. (d), where the symmetry of the system is consistent with the $+E$ case, so the corresponding component relationship remains unchanged, but both non-zero components $\chi_{xy}$ and $\chi_{yx}$have opposite signs.
}\label{fig3}
\end{figure}
\indent To identify the monolayers that realize this mechanism, we first exclude all layer groups corresponding to centrosymmetric point groups, as well as $LG_{74}$, $LG_{78}$, and $LG_{79}$. These layer groups either contain inversion symmetry $I$, or contain $M_z$ and $C_{nz}$ ($n>2$) simultaneously, which require all Edelstein tensor components to vanish identically. For a given initial layer group $LG_L$, all possible stacking operators $\hat{P}$ are constructed using quotient group theory, allowing the full bilayer symmetry group $LG_B$ to be determined. By comparing the tensor constraints before and after stacking, newly enforced zero components in the bilayer directly signal that the corresponding components in $L'$ are opposite to those in $L$. A complete enumeration of stacking configurations realizing this case is provided in Sec.~III of the SM\cite{sup}.
\paragraph{(ii) Electric-field-induced interlayer-opposite components.}
In the second scenario, more stringent symmetry constraints require that $\chi_{ij}^{B}$, $\chi_{ij}^{L}$ and $\chi_{ij}^{L'}$ vanish in the absence of external perturbations. In this case, an in-plane current cannot generate nonequilibrium spin polarizations on the two layers locally and supports only the symmetry-allowed charge response. Thus, in Type-II systems, the vertical electric field plays a dual role: it first acts as a symmetry-breaking field that is necessary to generate the Edelstein response, and once the response is activated, it further controls the relative contributions from the two layers.

This mechanism applies precisely to the layer groups excluded in case (i). Because the components vanish at zero field for each monolayer and corresponding bilayer, we must analyze the symmetry of the bilayer after applying the out-of-plane electric field. The field explicitly breaks all $LG_B^{\mathrm{inter}}$ operations while preserving $LG_B^{\mathrm{intra}}$ operations, thereby lifting symmetry constraints imposed by $LG_B^{\mathrm{inter}}$. Notably, acting with a $LG_B^{\mathrm{inter}}$ operation on a bilayer under an applied electric field is equivalent to reversing the direction of the electric field. Consequently, tensor components constrained by $LG_B^{\mathrm{inter}}$ symmetry reverse sign under field inversion, giving rise to opposite spin magnetizations on the two layers. For the LEE to occur, at least one tensor component must be unconstrained by $LG_B^{\mathrm{intra}}$ while being constrained by $LG_B^{\mathrm{inter}}$. A complete symmetry classification for this case is presented in Sec.~III of the SM\cite{sup}.

\indent This symmetry-based classification establishes a unified criterion framework for identifying nonmagnetic bilayers that host the LEE. By systematically enumerating stacking operators $\hat{P}$ and analyzing the resulting bilayer symmetries, one can determine whether Type-I or Type-II LEE are allowed without further invoking microscopic modeling. Collecting the results from both scenarios, we find that the realization of the LEE in stacked bilayers depends solely on the point-group symmetry of the bilayer, as summarized in Tab.~I. The presence of at least one $LG_B^{\mathrm{inter}}$ symmetry operation is essential, whereas inversion symmetry in the bilayer does not forbid and may even permit the effect. Remarkably, the LEE is therefore ubiquitous in nonmagnetic bilayers: any of the 80 layer-group materials can realize it through an appropriate stacking configuration. For example, inversion stacking applies naturally to noncentrosymmetric monolayers, while direct translation stacking suffices for centrosymmetric monolayers. General stacking operators for the layer groups and representative candidate materials for realizing the LEE are listed in Sec.~IV of the SM\cite{sup}.\\
\indent \textit{LEE in realistic materials}---We now apply the unified framework to semiconducting Janus MoSSe monolayer~\cite{MoSSe-1,MoSSe-2,MoSSe-3}. Since the relevant transport response is dominated by low-energy states near the $K$ valleys, we employ a fitted valley $\boldsymbol{k}\cdot\boldsymbol{p}$ model~\cite{kp}. The little group at $K$ is $C_3$, allowing cubic warping terms in the effective Hamiltonian. Including these terms in a spin-mixed model shows that they only weakly renormalize the response magnitude and do not affect the symmetry-dictated tensor form (see details in Sec.~V of SM\cite{sup}). Its layer group is $p3m1$ (No.~69), embedded in the hexagonal lattice group $D_{6h}$. The quotient construction then gives four inequivalent stacking operators, $\{\hat{E},\hat{2}_{001},\hat{I},\hat{M}_{001}\}$ (see details in Sec.~VI of the SM\cite{sup}). Among the four stacking operators, $\hat{I}$ and $\hat{M}_{001}$ yield Type-I LEE bilayers with point groups $D_{3d}$ and $D_{3h}$, respectively, whereas $\hat{E}$ and $\hat{2}_{001}$ give LEE-forbidden $C_{3v}$ bilayers~\cite{companion_paper}. The vertical electric field then reduces the Type-I LEE bilayer point group to the $C_{3v}$ subgroup, without changing the pristine allowed tensor components in constituent monolayers. Therefore, its effect is mainly to polarize the electronic states between the two layers and break the layer equivalence contributions\cite{magnetoelectric_couping2}. Such asymmetric modulation behavior enables the coupling among electric field polarity, the layer degree of freedom, and the induced spin polarization direction.\\
\indent First-principles calculations offer a realistic platform for validating the above symmetry analysis. Here we focus on the bilayer MoSSe stacked by $\hat{P} = \hat{I}$. As shown in Fig.~3(a), the optimized interlayer spacing $d=3.15~\AA$. It is worth noting that this stacking shifts the $\Gamma$-derived valence band closer to the Fermi level, whereas the K-valley preserve their monolayer-like spin-gapped character, as discussed in Sec.~VII of the SM\cite{sup}. Under a experimentally accessible finite out-of-plane electric field with $|E|=0.2 \text{ eV}/\AA$\cite{electric-experiment}, electrons near the Fermi level become localized predominately on one layer, as the layer-projected band structures shown in Fig.~3(c) and 3(d). Consequently, an in-plane current along the $x$ ($y$) direction induces a net spin magnetization along $-y$ ($x$) direction on the selected layer [Fig.~3(e) and 3(f)]. Reversing the electric field switches the dominant layer contribution and simultaneously reverses the layer-resolved spin magnetization. A quantitative discussion of the LEE magnitude and possible experimental observability is provided in Sec.~VIII of the SM\cite{sup}. Finite-temperature calculations further confirm that the predicted LEE is experimentally detectable (see Fig.~S3 of the SM~\cite{sup}). In Sec.~IX of the SM\cite{sup}, we also calculate the LEE of MoTe$_2$\cite{MoTe2-1,MoTe2-2,MoTe2-3}. The monolayer belongs to layer group $p\bar{6}$m2 and possesses the symmetries $C_{3z}$ and $M_z$, which enforce all tensor components to vanish; it therefore belongs to class~(ii). The bilayer is formed via direct stacking with the identity operation $E$, so the bilayer remains in layer group P$\bar{6}$m2. Under an external electric field, two non-zero tensor components, $\chi_{xy}$ and $\chi_{yx}$, are induced with opposing signs. Moreover, bilayer 1T'-WTe$_2$ provides a more direct metallic platform for investigating the LEE, as discussed in Sec.~X of the SM\cite{sup}.

\indent \textit{Discussion.}---In this Letter, we have introduced the LEE as the natural counterpart of the layer Hall effect within the broader Hall-family of transport phenomena. Our symmetry analysis demonstrates that this effect is not restricted to specific materials but is generically realizable in bilayer stacking systems. The two types of LEE are complementary in terms of potential device applications, as discussed in detail in Sec.~XI of the SM~\cite{sup}. For the semiconducting TMDs discussed above, the LEE can be experimentally probed using a dual-gated van der Waals device, as discussed in Sec.~IX of the SM~\cite{sup}. It is worth noting that the LEE and the layer-dependent spin Hall effect both arise from spin-layer coupling. Nevertheless, they are governed by distinct physical mechanisms and are therefore associated with different device functionalities, as discussed in detail in Sec.~XII of the SM\cite{sup}. Considering that there are various experimentally synthesized and theoretically predicted 2D non-magnetic materials with their applications\cite{experiment,C2DB,BiDB,LEEapplication1,LEEapplication2,LEEapplication3}, the LEE not only provides a wide configuration space for designing spin magnetization-switchable stacking bilayer spin-orbit logic devices\cite{spin-orbit-logic}, but also is unique in its potential to switch between two opposite spin magnetization configurations, thereby further enabling the manipulation of the "0" and "1" states in magnetic memory devices\cite{magnetoelectric_couping1,application1,application2}.\\
\indent \emph{Acknowledgments}--- This work is supported by the National Natural Science Foundation of China (Grant No. 12574070 and Grant No. 12504223), the China Postdoctoral Science Foundation (2025M773383 and GZC20252231), the China Postdoctoral Science Foundation - Hunan Joint Support Program (2025T002HN), the Excellent Youth Funding of Hunan Provincial Education Department (22B0175).\\
\indent Binchang Zhou and Pan Zhou contributed equally to this work.

\bibliography{references}
\end{document}